\documentclass[aps,prd,twocolumn,superscriptaddress,showpacs,preprintnumbers,amsmath,amssymb]{revtex4-1}
\usepackage{graphicx} 
\usepackage{dcolumn}  
\usepackage{rotating}
\usepackage{color}

\begin{document}



\title{ \quad\\[1.0cm]Update of inclusive cross sections of single and pairs of identified light charged hadrons}
\noaffiliation
\affiliation{University of the Basque Country UPV/EHU, 48080 Bilbao}
\affiliation{Beihang University, Beijing 100191}
\affiliation{Brookhaven National Laboratory, Upton, New York 11973}
\affiliation{Budker Institute of Nuclear Physics SB RAS, Novosibirsk 630090}
\affiliation{Faculty of Mathematics and Physics, Charles University, 121 16 Prague}
\affiliation{Chonnam National University, Gwangju 61186}
\affiliation{University of Cincinnati, Cincinnati, Ohio 45221}
\affiliation{Deutsches Elektronen--Synchrotron, 22607 Hamburg}
\affiliation{Duke University, Durham, North Carolina 27708}
\affiliation{University of Florida, Gainesville, Florida 32611}
\affiliation{Department of Physics, Fu Jen Catholic University, Taipei 24205}
\affiliation{Gifu University, Gifu 501-1193}
\affiliation{SOKENDAI (The Graduate University for Advanced Studies), Hayama 240-0193}
\affiliation{Department of Physics and Institute of Natural Sciences, Hanyang University, Seoul 04763}
\affiliation{University of Hawaii, Honolulu, Hawaii 96822}
\affiliation{High Energy Accelerator Research Organization (KEK), Tsukuba 305-0801}
\affiliation{J-PARC Branch, KEK Theory Center, High Energy Accelerator Research Organization (KEK), Tsukuba 305-0801}
\affiliation{Forschungszentrum J\"{u}lich, 52425 J\"{u}lich}
\affiliation{IKERBASQUE, Basque Foundation for Science, 48013 Bilbao}
\affiliation{Indian Institute of Science Education and Research Mohali, SAS Nagar, 140306}
\affiliation{Indian Institute of Technology Guwahati, Assam 781039}
\affiliation{Indian Institute of Technology Hyderabad, Telangana 502285}
\affiliation{Indian Institute of Technology Madras, Chennai 600036}
\affiliation{Indiana University, Bloomington, Indiana 47408}
\affiliation{Institute of High Energy Physics, Chinese Academy of Sciences, Beijing 100049}
\affiliation{Institute of High Energy Physics, Vienna 1050}
\affiliation{Institute for High Energy Physics, Protvino 142281}
\affiliation{INFN - Sezione di Napoli, 80126 Napoli}
\affiliation{INFN - Sezione di Torino, 10125 Torino}
\affiliation{Advanced Science Research Center, Japan Atomic Energy Agency, Naka 319-1195}
\affiliation{J. Stefan Institute, 1000 Ljubljana}
\affiliation{Institut f\"ur Experimentelle Teilchenphysik, Karlsruher Institut f\"ur Technologie, 76131 Karlsruhe}
\affiliation{King Abdulaziz City for Science and Technology, Riyadh 11442}
\affiliation{Korea Institute of Science and Technology Information, Daejeon 34141}
\affiliation{Korea University, Seoul 02841}
\affiliation{Kyungpook National University, Daegu 41566}
\affiliation{\'Ecole Polytechnique F\'ed\'erale de Lausanne (EPFL), Lausanne 1015}
\affiliation{P.N. Lebedev Physical Institute of the Russian Academy of Sciences, Moscow 119991}
\affiliation{Faculty of Mathematics and Physics, University of Ljubljana, 1000 Ljubljana}
\affiliation{Ludwig Maximilians University, 80539 Munich}
\affiliation{Luther College, Decorah, Iowa 52101}
\affiliation{University of Maribor, 2000 Maribor}
\affiliation{Max-Planck-Institut f\"ur Physik, 80805 M\"unchen}
\affiliation{School of Physics, University of Melbourne, Victoria 3010}
\affiliation{University of Mississippi, University, Mississippi 38677}
\affiliation{University of Miyazaki, Miyazaki 889-2192}
\affiliation{Moscow Physical Engineering Institute, Moscow 115409}
\affiliation{Moscow Institute of Physics and Technology, Moscow Region 141700}
\affiliation{Graduate School of Science, Nagoya University, Nagoya 464-8602}
\affiliation{Universit\`{a} di Napoli Federico II, 80055 Napoli}
\affiliation{Nara Women's University, Nara 630-8506}
\affiliation{National Central University, Chung-li 32054}
\affiliation{National United University, Miao Li 36003}
\affiliation{Department of Physics, National Taiwan University, Taipei 10617}
\affiliation{Nippon Dental University, Niigata 951-8580}
\affiliation{Niigata University, Niigata 950-2181}
\affiliation{Novosibirsk State University, Novosibirsk 630090}
\affiliation{Osaka City University, Osaka 558-8585}
\affiliation{Pacific Northwest National Laboratory, Richland, Washington 99352}
\affiliation{Peking University, Beijing 100871}
\affiliation{University of Pittsburgh, Pittsburgh, Pennsylvania 15260}
\affiliation{Theoretical Research Division, Nishina Center, RIKEN, Saitama 351-0198}
\affiliation{RIKEN BNL Research Center, Upton, New York 11973}
\affiliation{University of Science and Technology of China, Hefei 230026}
\affiliation{Showa Pharmaceutical University, Tokyo 194-8543}
\affiliation{Soochow University, Suzhou 215006}
\affiliation{Soongsil University, Seoul 06978}
\affiliation{Sungkyunkwan University, Suwon 16419}
\affiliation{Department of Physics, Faculty of Science, University of Tabuk, Tabuk 71451}
\affiliation{Tata Institute of Fundamental Research, Mumbai 400005}
\affiliation{Department of Physics, Technische Universit\"at M\"unchen, 85748 Garching}
\affiliation{School of Physics and Astronomy, Tel Aviv University, Tel Aviv 69978}
\affiliation{Toho University, Funabashi 274-8510}
\affiliation{Department of Physics, Tohoku University, Sendai 980-8578}
\affiliation{Earthquake Research Institute, University of Tokyo, Tokyo 113-0032}
\affiliation{Department of Physics, University of Tokyo, Tokyo 113-0033}
\affiliation{Tokyo Institute of Technology, Tokyo 152-8550}
\affiliation{Tokyo Metropolitan University, Tokyo 192-0397}
\affiliation{Virginia Polytechnic Institute and State University, Blacksburg, Virginia 24061}
\affiliation{Wayne State University, Detroit, Michigan 48202}
\affiliation{Yonsei University, Seoul 03722}
  \author{R.~Seidl}\affiliation{RIKEN BNL Research Center, Upton, New York 11973} 
  \author{I.~Adachi}\affiliation{High Energy Accelerator Research Organization (KEK), Tsukuba 305-0801}\affiliation{SOKENDAI (The Graduate University for Advanced Studies), Hayama 240-0193} 
  \author{H.~Aihara}\affiliation{Department of Physics, University of Tokyo, Tokyo 113-0033} 
  \author{D.~M.~Asner}\affiliation{Brookhaven National Laboratory, Upton, New York 11973} 
  \author{V.~Aulchenko}\affiliation{Budker Institute of Nuclear Physics SB RAS, Novosibirsk 630090}\affiliation{Novosibirsk State University, Novosibirsk 630090} 
 \author{T.~Aushev}\affiliation{Moscow Institute of Physics and Technology, Moscow Region 141700} 
  \author{I.~Badhrees}\affiliation{Department of Physics, Faculty of Science, University of Tabuk, Tabuk 71451}\affiliation{King Abdulaziz City for Science and Technology, Riyadh 11442} 
  \author{P.~Behera}\affiliation{Indian Institute of Technology Madras, Chennai 600036} 
  \author{K.~Belous}\affiliation{Institute for High Energy Physics, Protvino 142281} 
  \author{J.~Bennett}\affiliation{University of Mississippi, University, Mississippi 38677} 
% ...
\collaboration{The Belle Collaboration}

\noaffiliation
\begin{abstract}
We report new measurements of the production cross sections of pairs of charged pions and kaons as a function of their fractional energies using various fractional-energy definitions. Two different fractional-energy definitions were used and compared to the conventional fractional-energy definition reported previously. The new variables aim at either identifying dihadron cross sections in terms of single-hadron fragmentation functions, or to provide a means of characterizing the transverse momentum created in the fragmentation process. The results were obtained applying the updated initial-state radiation correction used in other recent Belle publications on light-hadron production cross sections. In addition, production cross sections of single charged pions, kaons, and protons were also updated using this initial-state radiation correction. The cross sections are obtained from a $558\,{\rm fb}^{-1}$ data sample collected at the $\Upsilon(4S)$ resonance with the Belle detector at the KEKB asymmetric-energy $e^+ e^-$
collider. 
\end{abstract}


\maketitle

\tighten

{\renewcommand{\thefootnote}{\fnsymbol{footnote}}}
\setcounter{footnote}{0}

The hadronization of highly energetic partons into final-state hadrons is often parameterized in terms of fragmentation functions. They are nonperturbative objects that at present cannot be calculated from first principles in the theory of the strong interaction, quantum chromodynamics (QCD). Factorization proofs, when applicable, allow one to extract fragmentation functions from experimental data of various high-energy processes~\cite{Collins:2011zzd}, such as lepton-nucleon scattering, hadron-hadron collision, or electron-positron annihilation. In turn they can then be used to study in more detail the partonic flavor and spin structure of the nucleon. Fragmentation functions are generally parameterized in terms of the initial parton flavor, detected hadron type, the energy or momentum fraction the detected hadron carries relative to the initial parton, as well as variables sensitive to parton spin or transverse momentum relative to the parton momentum direction. The clean initial state of the electron-positron annihilation process serves as the ideal tool to study fragmentation although the sensitivity to the parton flavor is limited. Detecting more than one hadron in the final state can partially overcome this limitation.

In the initial Belle publication \cite{Seidl:2015lla}, the dihadron cross sections in electron-positron annihilation, $e^+e^-\rightarrow h_1h_2X$, were measured as a function of the fractional energies $z_i=2E_{h,i}/\sqrt{s}$ of the two hadrons in various topologies. Here, $\sqrt{s}$ is the center-of-mass (c.m.) energy and $E_{h,i}$ the c.m. energy of hadron \(i\). Theorists brought to our attention two different fractional energy or momentum definitions: one that facilitates the interpretation of cross sections for pairs of nearly back-to-back hadrons in terms of single-hadron fragmentation functions \cite{Altarelli:1979kv}, the other serves to highlight the transverse momentum produced in the fragmentation process \cite{Mulders:2019mqo}. Moreover, in these alternative definitions, no additional thrust or hemisphere requirements are explicitly necessary since their definitions take the selection of back-to-back hadrons originating from two different partons into account directly via scalar products between the two hadron four-momenta. This feature in turn allows the interpretation of the cross sections even at higher orders of the strong coupling, which might not be possible in the conventional definition. 
The first alternative definition is in fact the oldest definition overall, already suggested in Ref.~\cite{Altarelli:1979kv}. The fractional energy of the first hadron is the same as the nominal definition, written in terms of four-vectors for hadrons $P_i$ and the virtual photon $q$ as
\begin{equation}
z_1 = \frac{2 P_1 \cdot q}{q^2} \quad, 
\end{equation}
where $q\cdot q = s$ is the squared four-momentum of the virtual photon. The fractional-energy definition for the second hadron differs in that it includes scalar products of the two hadron four-momenta:
\begin{equation}
z_2 = u =  \frac{P_1 \cdot P_2}{P_1 \cdot q} \quad. 
\end{equation}
It thus has a maximal contribution where the two hadrons are back-to-back and small values when the hadrons are found within the same hemisphere.
This set of fractional momenta will be referred to as the AEMP \cite{Altarelli:1979kv} definition in the following.

The other alternative fractional-energy definition is in part similar to the AEMP definition, but puts more stress on the masses of the hadrons, $M_{h1/2}$, and is motived to assess the transverse-momentum dependence of single-hadron fragmentation functions in the two-hadron system:
\begin{equation}
  z_1 = \left( P_1 \cdot P_2 - \frac{M_{h1}^2 M_{h2}^2}{P_1\cdot P_2}\right) \frac{1}{P_2 \cdot q - M_{h2}^2\frac{P_1 \cdot q}{P_1\cdot P_2}}\quad,
\end{equation}
and vice versa for $z_2$ when interchanging the indices of the first and second hadron.
This definition will be referred to as the Mulders-VanHulse, or MVH \cite{Mulders:2019mqo}, definition throughout this publication.\\

Both fractional-energy definitions will be compared to the conventional definition. They are similar to the conventional definition in that they can be seen as the fraction of the initial parton energy that a hadron carries. However, while the conventional and AEMP definitions are typical scaling variables limited to values between zero and one, the MVH fractional energy can exceed these limits, especially when the two hadrons are in the same hemisphere. For nearly back-to-back hadron pairs they are expected to behave similarly \cite{Mulders:2019mqo}, however, the distributions are expected to be shifted to lower $z$ due to the presence of non-zero transverse momentum. 

In addition to reporting these three fractional-energy definitions, all cross section measurements in this publication use an updated version of the initial-state radiation (ISR) correction. Unlike the previous publication \cite{Seidl:2015lla} for the dihadron fractional-energy dependence, an ISR correction is used that enables the direct applicability in global fits. This updated ISR correction is also applied to the previously published \cite{martin,Seidl:2015lla} single pion, kaon, and proton cross sections $e^+e^-\rightarrow hX$ as a function of $z$, where the fractional energy is given by $z=2E_{h}/\sqrt{s}$. For the dihadron cross sections, also the ordering by hemisphere was removed, combining all hadron pair permutations of the same physics content, such as $\pi^+\pi^-$ and $\pi^-\pi^+$. 
Furthermore, all systematic uncertainties are now separated into components that are correlated over kinematic bins and those that ar uncorrelated.


The paper is organized as follows: In Section I the updated correction procedure will be discussed and consequently applied to obtain the updated single-hadron cross sections. In Section II the same update is performed also for the dihadron cross sections together with the combination of permutations.
In Section III the dihadron cross sections are compared to the new fractional-energy definitions before a summary concludes this publication.

\section{Update of single-hadron cross section measurements}
\begin{figure*}[t]
\begin{center}
\includegraphics[width=0.8\textwidth]{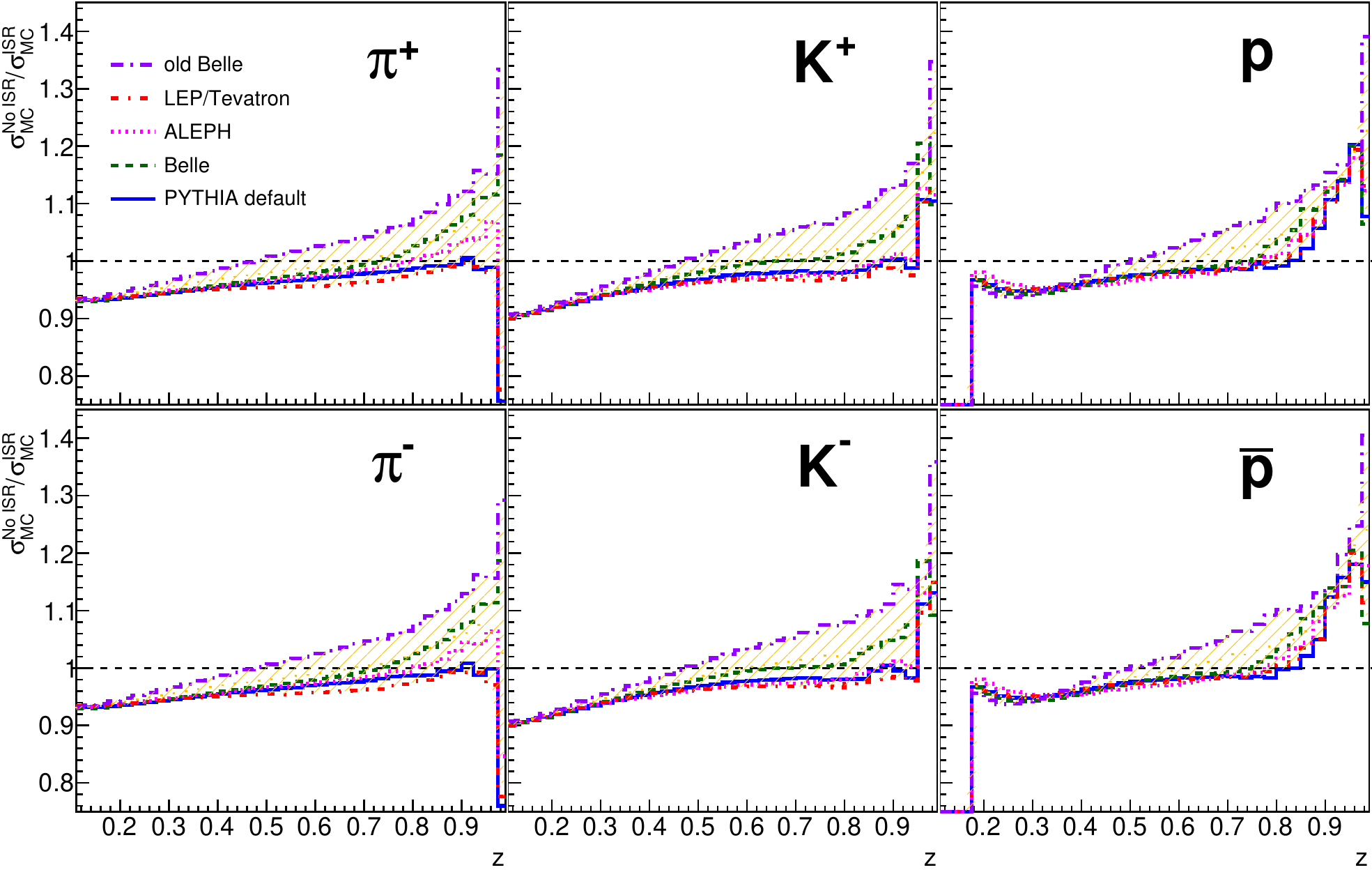}
\caption{\label{fig:isrzfraction}Non-ISR over ISR cross section ratios as a function of $z$ for pions, kaons, and protons for various MC tunes. The yellow, hatched regions display the variation of these ratios with tunes and are assigned as systematic uncertainties around the Belle tune.}
\end{center}
\end{figure*}
The analysis in this publication closely follows all steps mentioned in Ref.~\cite{Seidl:2015lla}. To recall that analysis, it is briefly described here. A total data set of 558 fb$^{-1}$ collected with the Belle detector at the center-of-mass (c.m.) energy of $\sqrt{s}=$ 10.58 GeV was used. Based on various detector components, charged tracks are initially identified as pions, kaons, protons, as well as electrons and muons. Hadron yields are then calculated in bins of fractional energy $z$ for each hadron. 36 equidistant bins between 0.1 and 1.0 are initially populated for pions, kaons, and protons in the single-hadron analysis. All yields are corrected for particle misidentification. Backgrounds from $\Upsilon(4S)$ decays, $\tau$ pair production, and two-photon processes are removed as detailed in previous publications \cite{Seidl:2017qhp,Seidl:2015lla}. Taking into account that the initial yields were extracted in the barrel part of the detector, acceptance and reconstruction efficiencies are corrected for as the next step of the analysis. The variation of the acceptance effects based on several fragmentation tunes in {\sc pythia} \cite{Sjostrand:2006za} are assigned as systematic uncertainties. Weak decays are removed based on Monte Carlo (MC) information in the next step. Due to different preferences by global fitting groups, both cross sections, with and without weak decays, will be provided in the Supplemental Material \cite{supplement}. The last step in the correction chain is the ISR correction. This correction is the main difference to the previous publications \cite{martin,Seidl:2015lla} and is therefore explained in more detail in the following.
\subsection{Updated ISR correction}
The previously published cross sections of Refs.~\cite{martin,Seidl:2015lla} utilized an ISR correction that was ultimately not quite as usable to the global fitters. It uses an arbitrary value for the actual c.m.~energy of the quark-antiquark system, based on MC, and keeps only the event fraction above that value. In the previous publications an energy above 99.5\% of the nominal c.m.~energy was selected. The problem with this selection is that global fitters need to implement a similar selection in order to use these data. In contrast, the updated ISR correction should be directly applicable. The single-hadron cross sections as a function of $z$ using the updated ISR correction approach are presented here in order to provide more practical input for global analyses. The ISR correction is obtained by calculating the ratios of MC cross sections without ISR over those with ISR included in the simulation (corresponding to the setting MSTP(11) to be zero or one in {\sc pythia}). These ISR ratios are shown in Fig.~\ref{fig:isrzfraction} for single hadrons, where various {\sc pythia} MC tunes are shown for comparison. The explicit differences in the {\sc pythia} settings are tabulated in the Supplemental Material \cite{supplement}. The variation due to these different tunes is assigned as a systematic uncertainty around the Belle tune; however, unlike the case for previous measurements \cite{Seidl:2017qhp,Seidl:2019jei}, where large effects occur in the tails of the distributions of hadron mass or transverse momenta, the overall effect here is only about 10\% around unity and the variations between tunes are even smaller. One can see that at small fractional energies the yields are larger when including ISR, while at higher fractional energies ISR effects reduce the phase space and the ratios exceed unity. Generally, also the variations between tunes increase with higher fractional energies.  
\subsection{Systematics and results}
Unlike the previous fractional-energy dependent single-hadron measurement \cite{martin,Seidl:2015lla}, systematic uncertainties are now separated into correlated and uncorrelated uncertainties. The uncorrelated uncertainties are generally related to the statistical uncertainties in the MC samples used to extract each correction, while the correlated uncertainties correspond to the variation of correction methods or MC tunes that affect all fractional-energy bins in a similar way. The correlated and uncorrelated systematic uncertainties are provided separately in the Supplemental Material where all correlated uncertainties and all uncorrelated uncertainties are added in quadrature. All single-hadron results are dominated by systematic uncertainties. These, in turn, are mostly dominated by the correlated systematic uncertainties from the tune variations in ISR, acceptance, and weak decay corrections at intermediate to high $z$. The uncertainties from these three corrections have been assigned together. Correlated uncertainties due to the non-$q\bar{q}$ background are especially larger at low fractional energies for all hadron types. PID uncertainties are also large at low $z$, and they are the dominant source of correlated systematics at large $z$ for both kaons and protons. At very high $z$, for kaons and protons, and at low $z$, the uncorrelated uncertainties are also sizable.
\begin{figure}[tb]
\begin{center}
\includegraphics[width=0.49\textwidth]{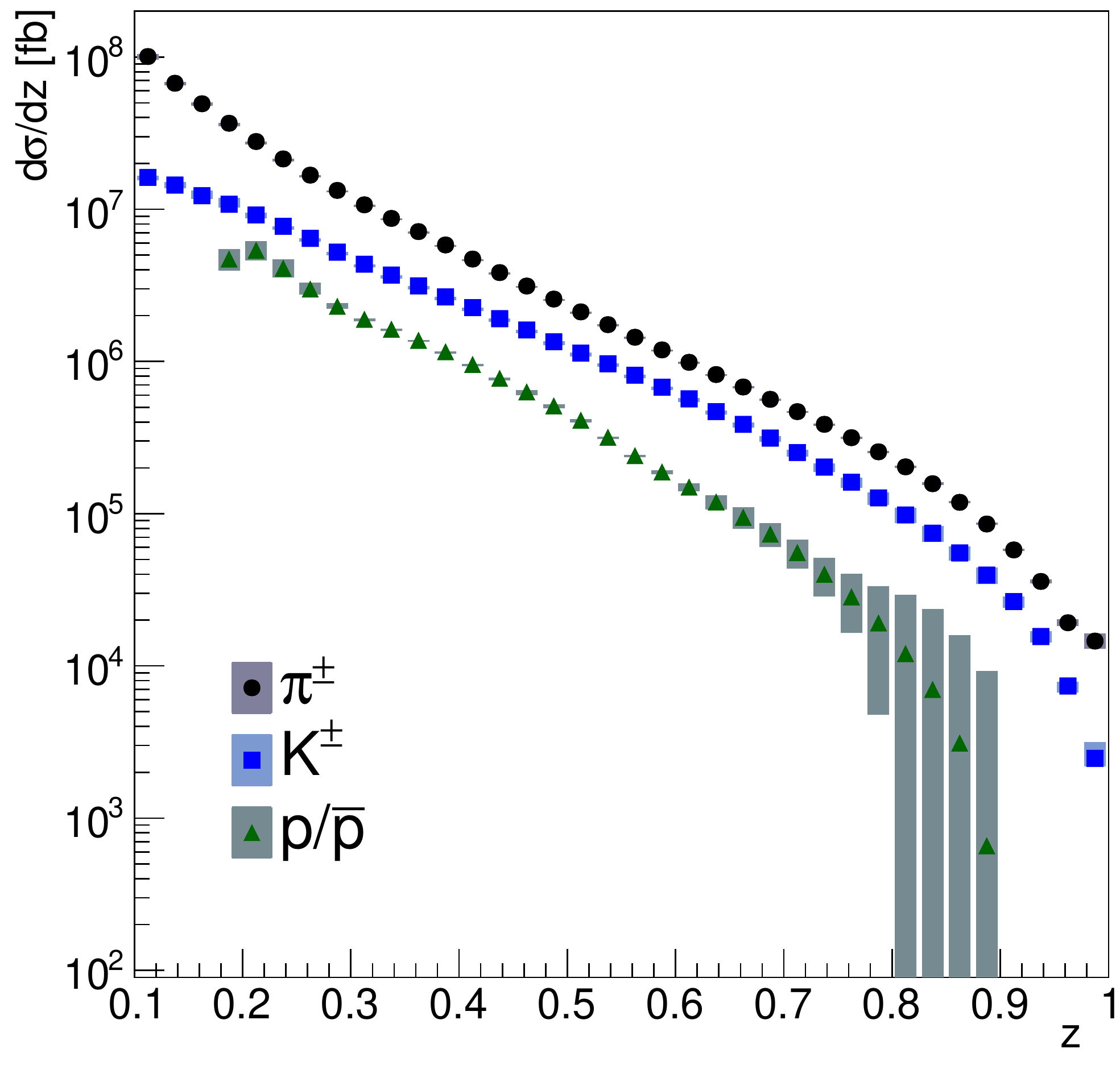}

\caption{\label{fig:allxsec_pid21}Differential cross sections for pions (black circles), kaons (blue squares), and protons (green triangles) as a function of  $z$ without any thrust requirement. The error boxes represent the systematic and error bars the statistical uncertainties.}
\end{center}
\end{figure}

The single-hadron cross sections using the updated ISR correction are presented in Fig.~\ref{fig:allxsec_pid21}. As the current ISR corrections are close to unity, the resulting cross sections are higher than the previously published ones \cite{Seidl:2015lla}. Previously, only about 60\% of the events were kept since the rest had a c.m. energy of the quark-antiquark system reduced by more than 0.5\% from the nominal c.m. energy. The general ordering of the pion, kaon, and proton cross sections does not change with this update. Pions are the lightest hadrons and are most abundant, especially at low $z$. At higher $z$ the shapes of pions and kaons are similar, which may be due to the favored fragmentation contribution ($u,d\rightarrow \pi$ and $s \rightarrow K$) or that the differences in quark and hadron masses are relevant only at smaller values of $z$.   
 We hope that these new measurements will be taken up for the next round of updates of the various fragmentation-function fitting groups \cite{deFlorian:2017lwf,deFlorian:2014xna,Bertone:2018ecm,Sato:2016wqj}. With the separation of systematic uncertainties in correlated and uncorrelated contributions, the significance of these results will be increased.

\section{Update of dihadron cross section measurements}

\begin{figure*}[t]
\begin{center}
\includegraphics[width=0.9\textwidth]{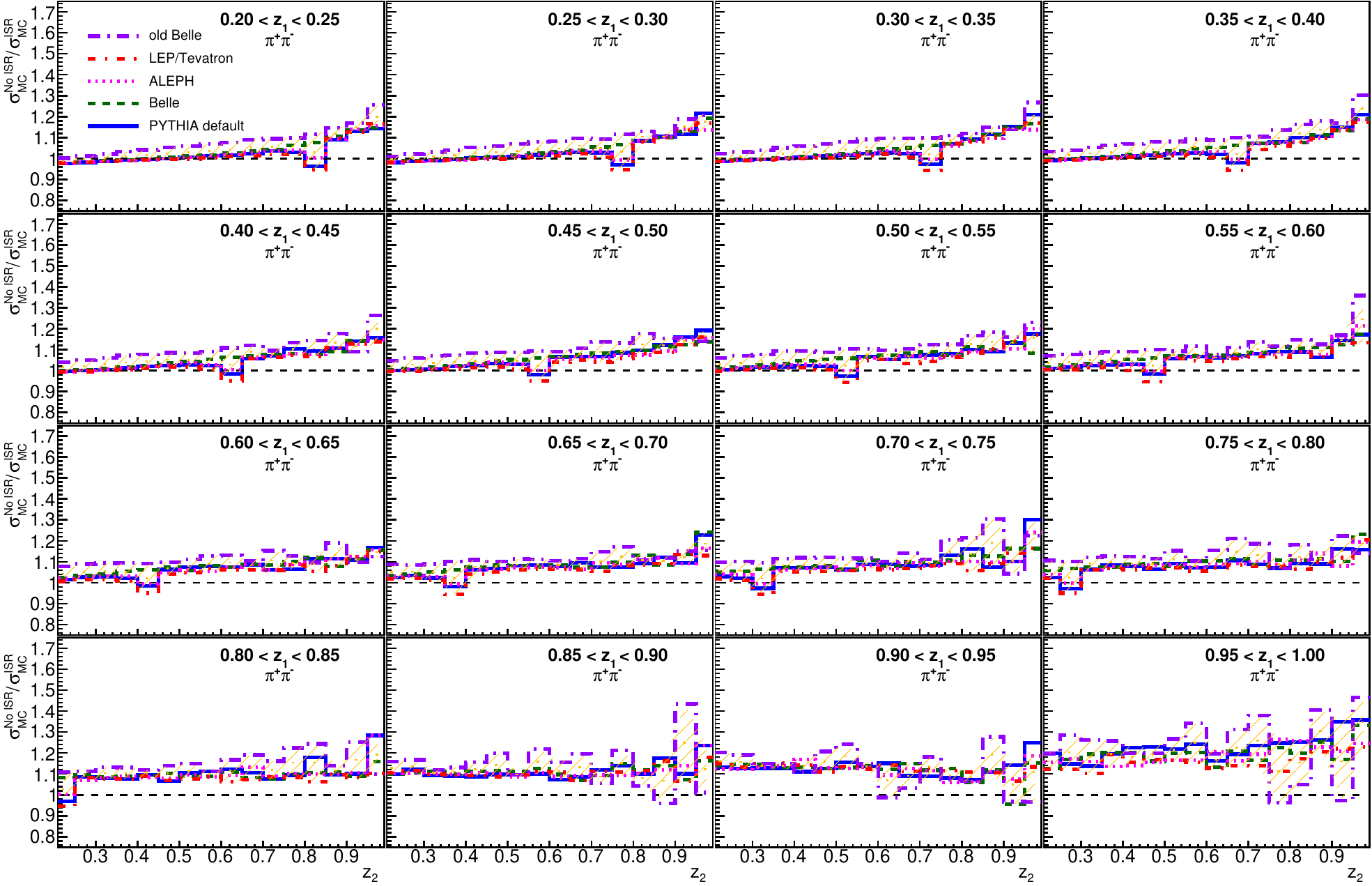}
\caption{\label{fig:isrzfractionpair}Non-ISR over ISR cross section ratios as a function of $z_2$ in bins of $z_1$ using the conventional fractional energy definitions for opposite-sign pion pairs without hemisphere restriction for various MC tunes (as labeled). The yellow, hatched regions display the variation of these ratios with tunes and are assigned as systematic uncertainties.}
\end{center}
\end{figure*}
In the dihadron analysis, 16 equidistant fractional-energy bins each between 0.2 and 1.0 are chosen for all six particle-charge combinations of pions and kaons, where we combine the charge-conjugate combinations, and each fractional-energy definition. In the case of the dihadron analysis, the yields are also classified as to whether the hadrons are in the same or opposite hemisphere in addition to any topology, where the hemispheres are defined by the plane perpendicular to the thrust axis. Also a minimum thrust value $T>0.8$ was required for hemisphere-separated dihadrons. The event-shape variable thrust is calculated by maximizing the sum over all reconstructed charged particles and neutral clusters in an event by
\begin{equation}
  T \stackrel{max}{=} \frac{\sum_h |\mathbf{p}_h \cdot \hat{\mathbf{n}}|}{\sum_h |\mathbf{p}_h|}\quad ,
\end{equation}
where $\hat{\mathbf{n}}$ defines the thrust axis and also the hemispheres.
The analysis follows the same correction steps mentioned in the previous publication and mostly the ISR correction is performed differently to the measurement it supersedes \cite{Seidl:2015lla}. Also, here the ratios between MC yields with ISR switched off and on are taken as the basis for the ISR correction, while the variation of these ratios based on various MC tunes is taken as a correlated systematic uncertainty. The corresponding ISR ratios are displayed in Fig.~\ref{fig:isrzfractionpair} for $\pi^+\pi^-$ pairs without hemisphere restriction. As can be seen, in this case the correction factors are again moderate. For dihadrons in the same hemisphere, the corrections become larger at the kinematic edges where the different boosts for ISR events migrate pairs from opposite hemispheres into the same hemisphere. The variations between tunes, assigned as systematic uncertainties, are moderate for opposite-hemisphere and any-hemisphere dihadrons, while they get larger when the ratios themselves increase for same-hemisphere dihadrons in the tails of the distributions. In general, the inclusion of the {\sc pythia}-tune dependence of ISR and acceptance corrections leads to increased systematics compared to the previous results, albeit the impact of those being partially weakened by their correlation between $z$ bins.

\begin{figure*}[htb]
\begin{center}
\includegraphics[width=0.9\textwidth]{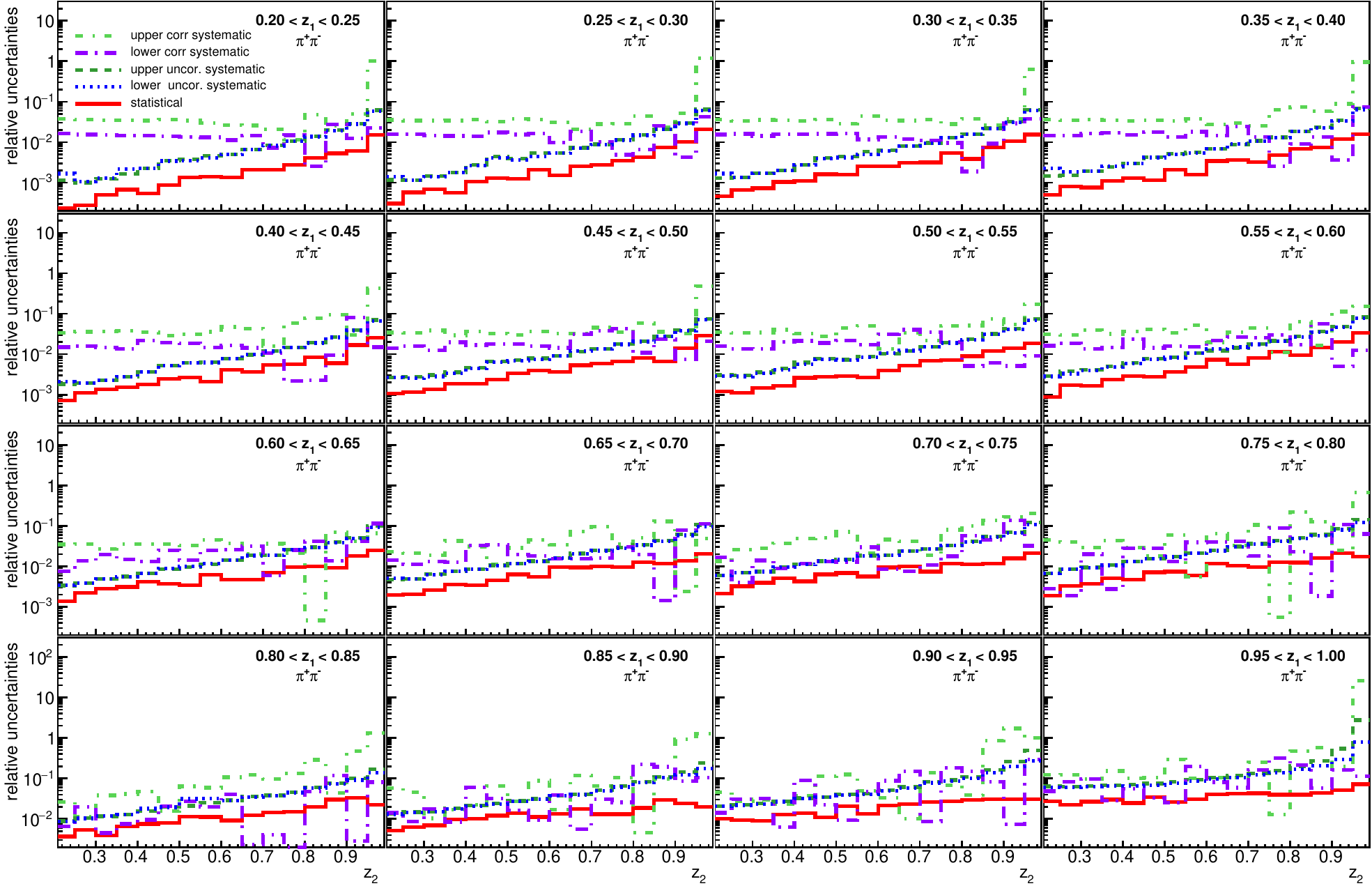}

\caption{\label{fig:systmix4}Relative statistical and systematic uncertainties as a function of  $z_2$ in bins of $z_1$ using the conventional fractional energy definitions for opposite-sign pion pairs without hemisphere restriction.}
\end{center}
\end{figure*}

In contrast to the results in the previous dihadron publication, the hadron permutations with same physics content, as well as the arbitrary ordering into first and second particle, have been combined after they have been confirmed to be consistent with each other. 
The final relative uncertainty budgets are shown in Fig.~\ref{fig:systmix4} for opposite-sign pion pairs without hemisphere restriction for the nominal fractional-energy definitions. As is the case for single hadrons, the dihadron measurements are systematics dominated. Correlated systematic uncertainties are predominantly larger than the uncorrelated uncertainties, except for very high fractional energies where both correlated and uncorrelated systematics become of similar size. The three largest contributions in the systematic uncertainties originate in the uncertainties in acceptance, weak decay and ISR corrections due to different MC tunes. These three sources of uncertainty are correlated among themselves and have therefore been evaluated as combined tune-dependence systematics. At lower $z$ the uncertainties due to the non-$q\bar{q}$ backgrounds are the largest, while at large $z$ systematics due to PID corrections are sizable, especially for pairs involving kaons.

\begin{figure*}[htb]
\begin{center}
\includegraphics[width=0.83\textwidth]{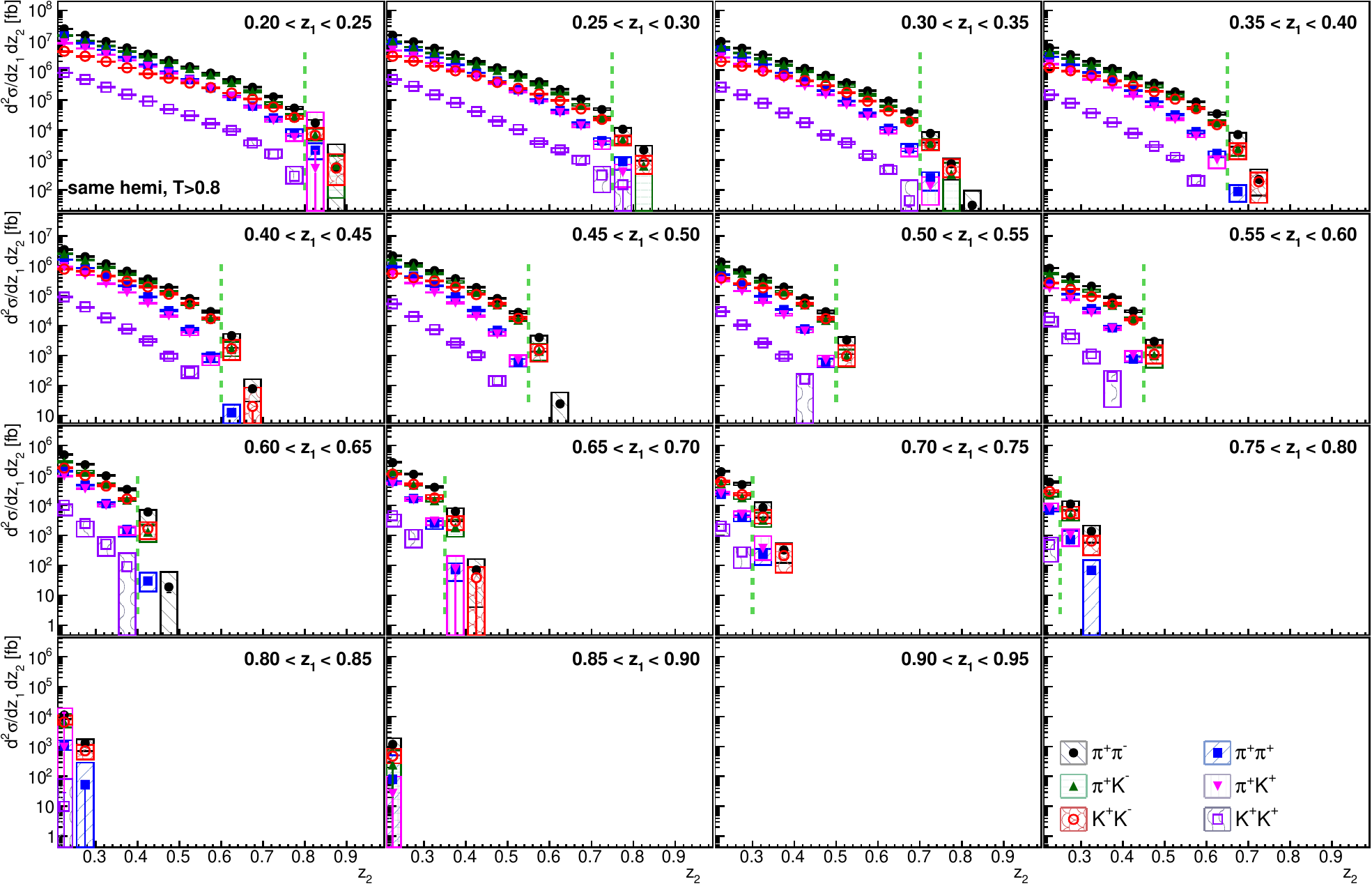}

\caption{\label{fig:figsame}Differential cross sections for $\pi^+\pi^-$ (black circles), $\pi^+\pi^+$ (blue squares), $\pi^+K^-$(green triangles), $\pi^+K^+$(magenta triangles), $K^+K^-$(red circles), and $K^+K^+$(purple squares) and their charge-conjugate states, as a function of  $z_2$ in bins of $z_1$ using the conventional fractional-energy definitions for dihadrons in the same hemisphere and using a thrust selection of $T>0.8$. The error boxes represent the combined systematic, and the error bars statistical, uncertainties. The green, dotted line represents the kinematic cutoff where the sum of the fractional energies exceeds unity.}
\end{center}
\end{figure*}

\begin{figure*}[htb]
\begin{center}
\includegraphics[width=0.83\textwidth]{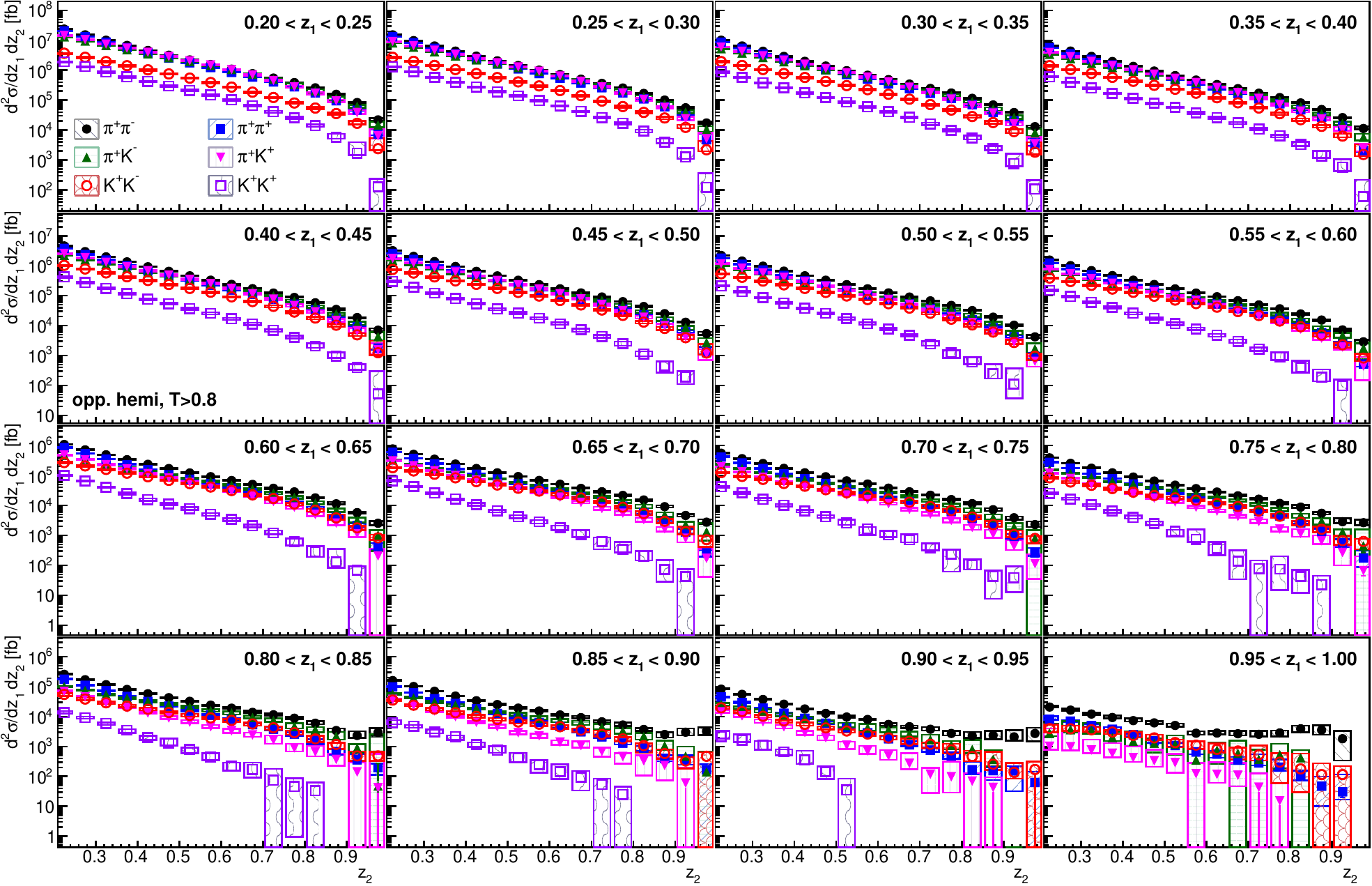}

\caption{\label{fig:figopp}Differential cross sections for $\pi^+\pi^-$ (black circles), $\pi^+\pi^+$ (blue squares), $\pi^+K^-$(green triangles), $\pi^+K^+$(magenta triangles), $K^+K^-$(red circles), and $K^+K^+$(purple squares) and their charge-conjugate states, as a function of  $z_2$ in bins of $z_1$ using the conventional fractional-energy definitions for dihadrons in opposite hemispheres and using a thrust selection of $T>0.8$. The error boxes represent the systematic, and the error bars statistical, uncertainties.}
\end{center}
\end{figure*}

\begin{figure*}[htb]
\begin{center}
\includegraphics[width=0.83\textwidth]{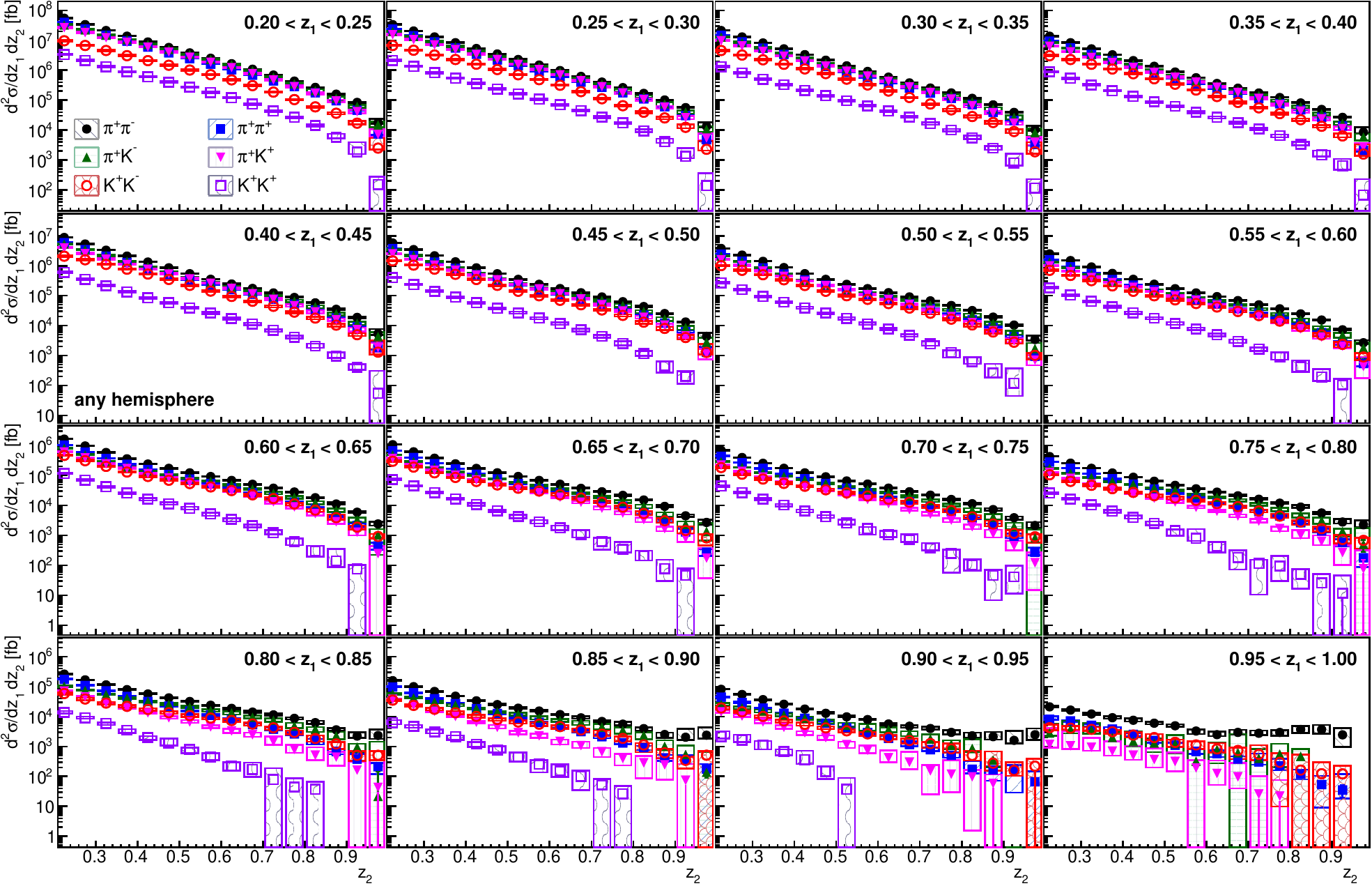}

\caption{\label{fig:figany}Differential cross sections for $\pi^+\pi^-$ (black circles), $\pi^+\pi^+$ (blue squares), $\pi^+K^-$(green triangles), $\pi^+K^+$(magenta triangles), $K^+K^-$(red circles), and $K^+K^+$(purple squares) and their charge-conjugate states, as a function of  $z_2$ in bins of $z_1$ using the conventional fractional-energy definitions for dihadrons without hemisphere or thrust selection. The error boxes represent the systematic, and the error bars statistical, uncertainties.}
\end{center}
\end{figure*}

The updated results for the dihadron cross sections as a function of the fractional energies are presented in Figs.~\ref{fig:figsame} to \ref{fig:figany} including charge-conjugate final states. Figure \ref{fig:figsame} displays the differential cross sections for dihadrons in the same hemisphere. As previously noted, both hadrons likely emerged from the same initial parton and as such, the sum of their fractional energies is bounded by unity. Same-sign pairs of any hadron type are generally more suppressed than opposite-sign pairs. Pions are generally slightly favored over kaons, and same-sign kaons are strongly suppressed. In the latter case strangeness has to be generated in the fragmentation process, while single kaons can originate from the initial strange or charmed partons.

When looking at dihadrons in opposite hemispheres, as shown in Fig.~\ref{fig:figopp}, pion pairs as well as pion-kaon combinations all have similar cross sections at small fractional energies and only at higher fractional energies opposite-sign pion pairs start to dominate. Similarly, opposite-sign kaon pairs, while suppressed at small fractional energies, have the second-highest cross sections at large fractional energies and opposite-sign pion-kaon combinations are of comparable magnitude.
It is interesting to note that in opposite hemispheres, same-sign pion-kaon pairs in the conventional definitions exceed the opposite sign pairs when both of the fractional energies are not too large. This behavior can be traced to charm decays producing more same-sign pion-kaon pairs. When weak decays are removed, the opposite pion-kaon pairs are again larger.
  
The cross sections without hemisphere restriction follow the opposite-hemisphere dihadrons at higher $z$ where only those can contribute via single-hadron fragmentation. At lower fractional energies the cross sections increase due to the contributions from same-hemisphere dihadrons.

\section{Comparison of fractional-energy/momentum definitions}
The two alternative fractional-energy definitions have been analyzed in the same way, following the same correction steps and using also a binning of 16 fractional-energy/momentum bins between 0.2 and 1.0. Generally, the behavior in all correction steps is quite similar for opposite-hemisphere and any-hemisphere dihadrons. For same-hemisphere dihadrons, hardly any events get selected in the first place due to the fact that the scalar products produce fractional energies below the minimum bin boundaries. This is consistent with the focus of these variables on nearly back-to-back geometries. Consequently, explicit same-hemisphere dihadrons are no longer considered for these fractional-energy definitions and only opposite-hemisphere and any-hemisphere dihadron combinations will be discussed.     
The overall corrections, as well as the resulting systematic uncertainties for the two other fractional-energy definitions, are again similar to those using the conventional definitions. In all cases the systematic uncertainties dominate over the statistical uncertainties and the same correction steps (predominantly ISR and acceptance corrections) provide the largest contributions to the systematic uncertainty budget.  

\begin{figure*}[htb]
\begin{center}
  \includegraphics[width=0.85\textwidth]{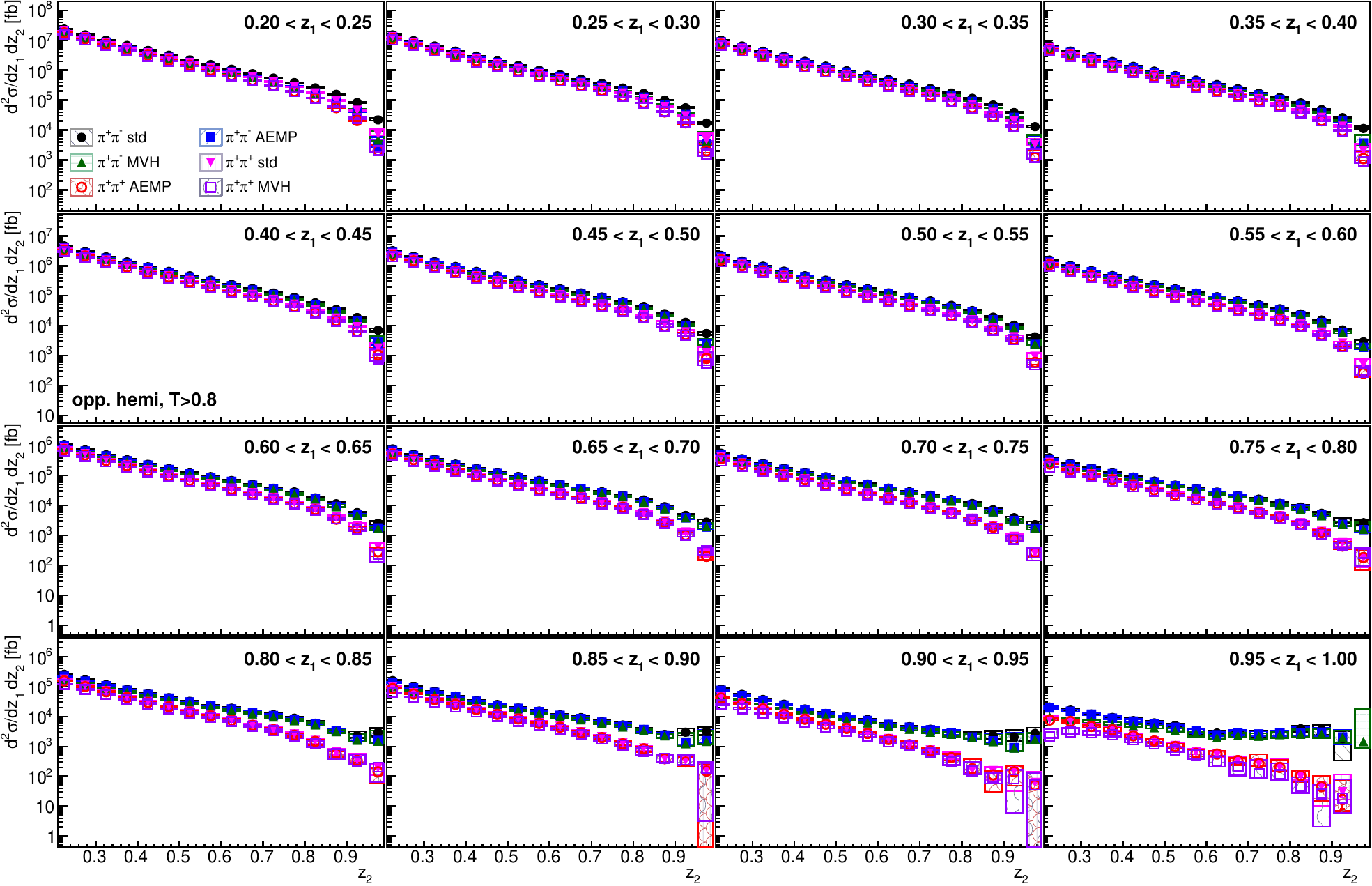}
  \includegraphics[width=0.85\textwidth]{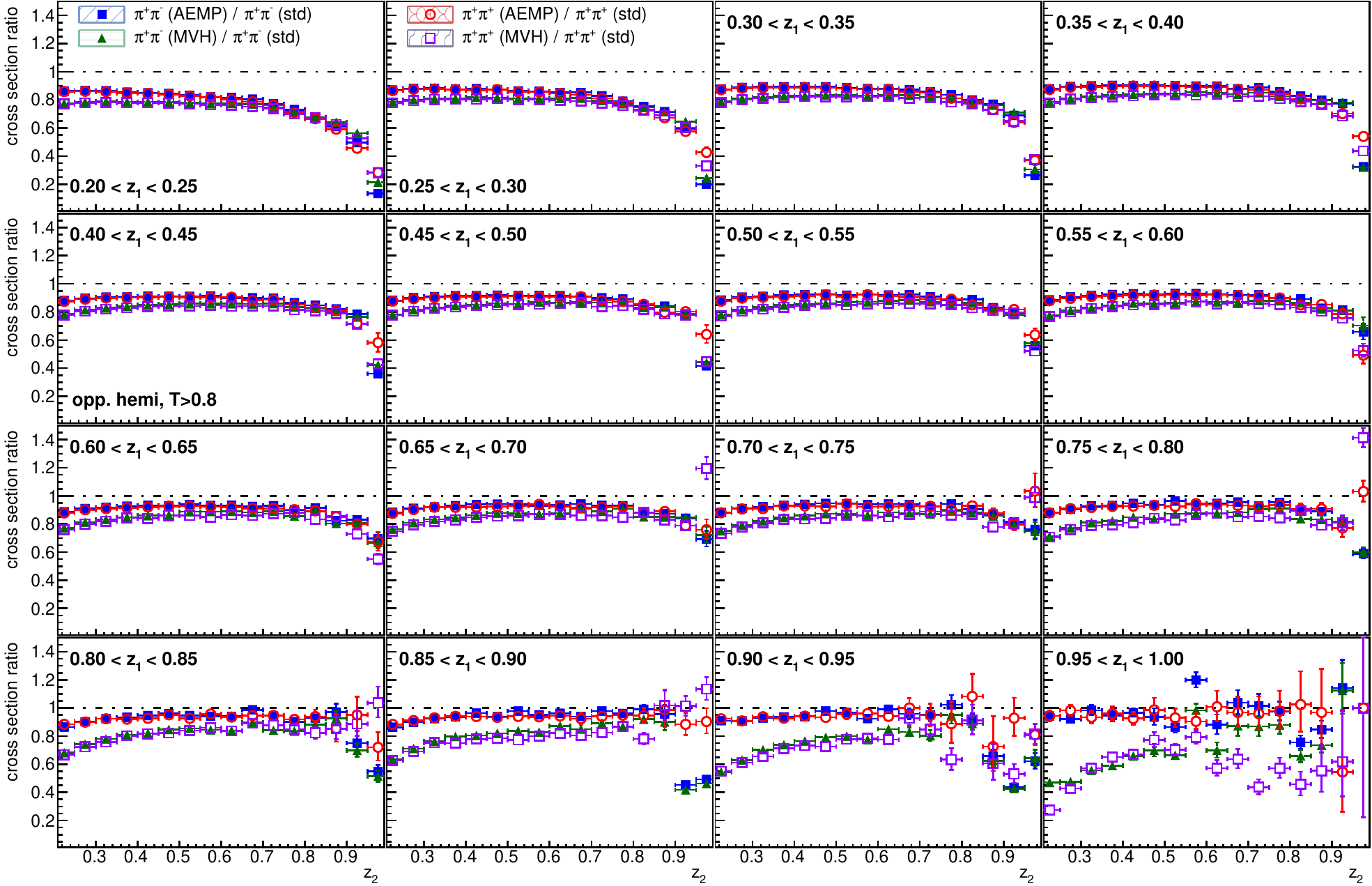}
\caption{\label{fig:zxsec_3_mix4_z7} Top: Differential cross sections for $\pi^+\pi^-$ and $\pi^+\pi^+$ pairs in opposite hemispheres as a function of $z_2$ in bins of $z_1$. The conventional $z$ definitions are displayed in black points and magenta triangles, respectively. Similarly the AEMP definitions are displayed by blue squares and red circles and the MVH definitions are displayed in green triangles and purple squares. The error boxes represent the systematic, and error bars the statistical, uncertainties. Bottom: Ratios of the pion pair cross sections for the alternative definitions to the corresponding ones using the conventional definitions. For better visibility, no systematic uncertainties are drawn.}
\end{center}
\end{figure*}

\begin{figure*}[htb]
  \begin{center}
    
  \includegraphics[width=0.85\textwidth]{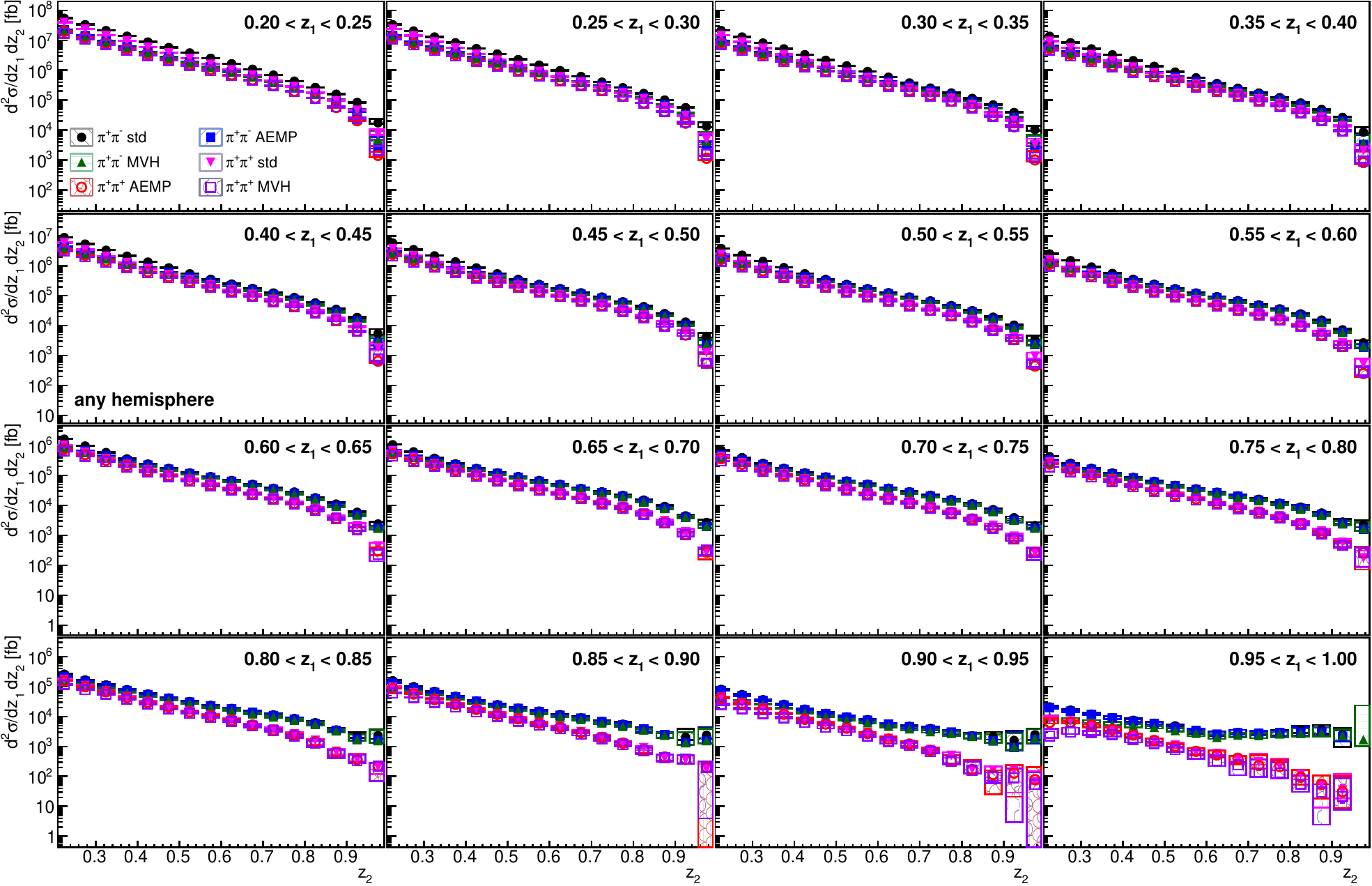}
  \includegraphics[width=0.85\textwidth]{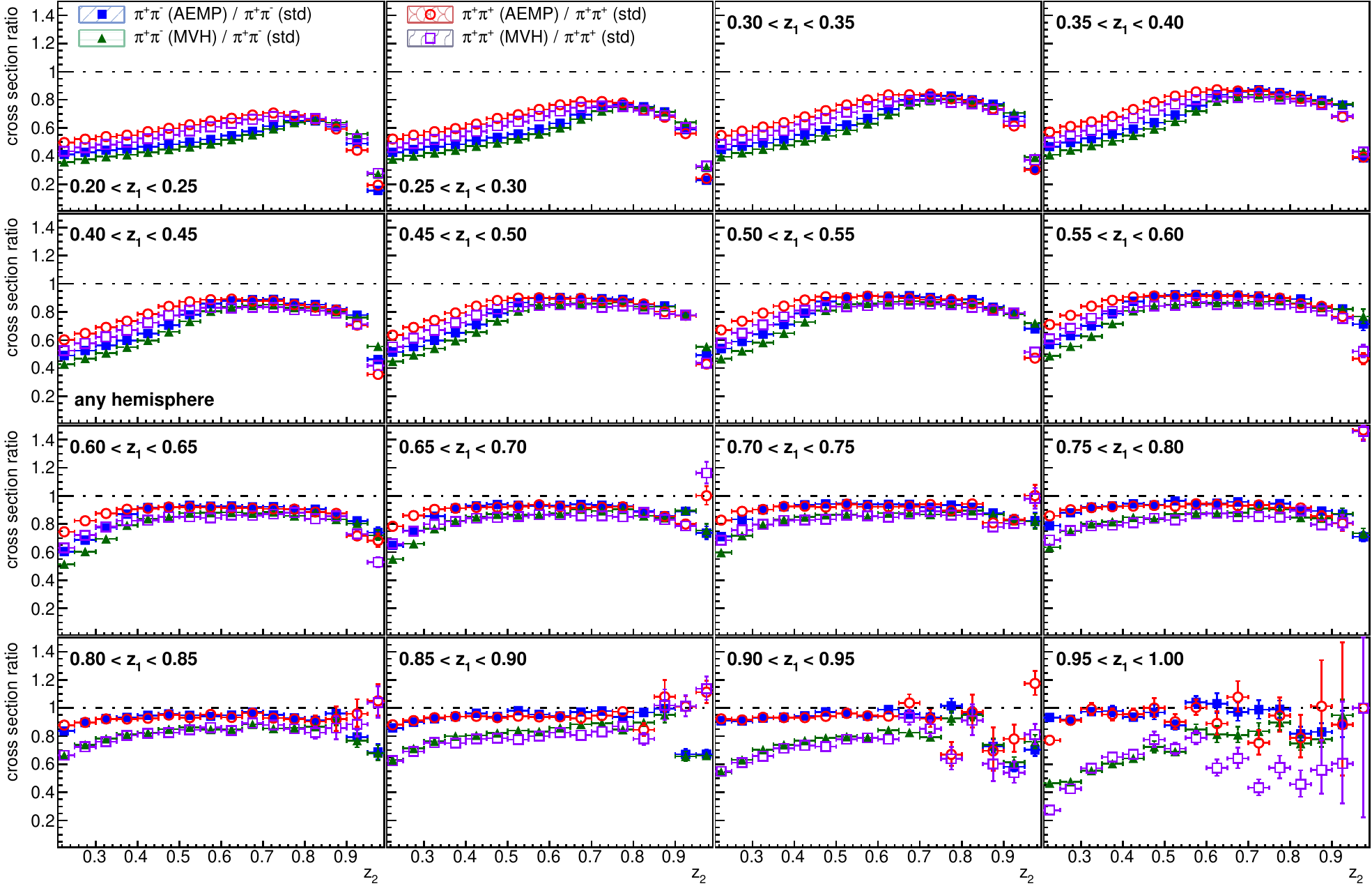}

\caption{\label{fig:zxsec_3_mix16_z7} Top: Differential cross sections for $\pi^+\pi^-$ and $\pi^+\pi^+$ pairs without hemisphere requirement as a function of $z_2$ in bins of $z_1$. The conventional $z$ definitions are displayed in black points and magenta triangles, respectively. Similarly the AEMP definitions are displayed by blue squares and red circles and the MVH definitions are displayed in green triangles and purple squares. The error boxes represent the systematic, and error bars the statistical, uncertainties. Bottom: Ratios of the pion pair cross sections for the alternative definitions to the corresponding ones using the conventional definitions. For better visibility, no systematic uncertainties are drawn.}
\end{center}
\end{figure*}

When comparing the different fractional-energy definitions for opposite-hemisphere dihadrons in Fig.~\ref{fig:zxsec_3_mix4_z7}, one sees that the cross sections are quite similar although both alternative definitions are slightly smaller. As the AEMP definitions are not symmetric, with the first hadron definition the same as the conventional definition and only the second hadron containing the dot-product, one can see that at high $z_1$ the cross sections for both definitions are very similar for all \(z_2\), while at low \(z_1\), they are quite different for basically any \(z_2\), and even more so at high \(z_2\).
The cross sections using the MVH definitions follow those of the AEMP definitions at small fractional energies, but are overall smaller and stay substantially smaller for higher $z_1$ and only approach the other definitions when both fractional energies become large. This is the expected behavior as the effect of nonzero transverse momentum would shift the fractional energies toward lower values, and only at the highest fractional energies the phase space for transverse momentum vanishes.
In Fig.~\ref{fig:zxsec_3_mix16_z7}, dihadron cross sections without hemisphere assignment are compared. Due to the scalar product between the two hadron momenta in the alternative definitions, there is hardly any difference between opposite-hemisphere cross sections and those not relying on a hemisphere assignment. The contributions from same-hemisphere dihadrons stick out at low fractional energies for the conventional definitions, which would appear below the fractional energies limit imposed here for the alternative definitions. 

For pion-kaon pairs (see Supplement Material), the addition of the actual hadron masses in the MVH definition results in a further suppression of the cross sections when $z_1$ is small and at moderate to larger values of $z_2$. Otherwise, the qualitative behavior is the same as for pions, where eventually at high fractional energies all definitions become comparable. 

\section{Summary}
In this paper, the single-hadron cross sections for charged pions, kaons, and protons, as well as the dihadron cross sections for pairs of charged pions and/or kaons in electron-positron annihilation were presented. In contrast to the previous publication of these cross sections \cite{Seidl:2015lla}, an updated ISR correction procedure was applied and systematic uncertainties were separated into uncorrelated and correlated contributions. These new results supersede the previous ones and should be used henceforth in global fits. Additionally, the dihadron cross sections for two alternative fractional-energy definitions were extracted. They behave similarly to those of the conventional definitions for opposite-hemisphere dihadrons at high fractional energies but allow a more direct single-hadron fragmentation interpretation even without hemisphere assignments.

\begin{acknowledgments}
  
We thank the KEKB group for the excellent operation of the
accelerator; the KEK cryogenics group for the efficient
operation of the solenoid; and the KEK computer group, and the Pacific Northwest National
Laboratory (PNNL) Environmental Molecular Sciences Laboratory (EMSL)
computing group for strong computing support; and the National
Institute of Informatics, and Science Information NETwork 5 (SINET5) for
valuable network support.  We acknowledge support from
the Ministry of Education, Culture, Sports, Science, and
Technology (MEXT) of Japan, the Japan Society for the 
Promotion of Science (JSPS), and the Tau-Lepton Physics 
Research Center of Nagoya University; 
the Australian Research Council including grants
DP180102629, 
DP170102389, 
DP170102204, 
DP150103061, 
FT130100303; 
Austrian Science Fund under Grant No.~P 26794-N20;
the National Natural Science Foundation of China under Contracts
No.~11435013,  
No.~11475187,  
No.~11521505,  
No.~11575017,  
No.~11675166,  
No.~11705209;  
Key Research Program of Frontier Sciences, Chinese Academy of Sciences (CAS), Grant No.~QYZDJ-SSW-SLH011; 
the  CAS Center for Excellence in Particle Physics (CCEPP); 
the Shanghai Pujiang Program under Grant No.~18PJ1401000;  
the Ministry of Education, Youth and Sports of the Czech
Republic under Contract No.~LTT17020;
the Carl Zeiss Foundation, the Deutsche Forschungsgemeinschaft, the
Excellence Cluster Universe, and the VolkswagenStiftung;
the Department of Science and Technology of India; 
the Istituto Nazionale di Fisica Nucleare of Italy; 
National Research Foundation (NRF) of Korea Grants
No.~2015H1A2A1033649, No.~2016R1D1A1B01010135, No.~2016K1A3A7A09005
603, No.~2016R1D1A1B02012900, No.~2018R1A2B3003 643,
No.~2018R1A6A1A06024970, No.~2018R1D1 A1B07047294; Radiation Science Research Institute, Foreign Large-size Research Facility Application Supporting project, the Global Science Experimental Data Hub Center of the Korea Institute of Science and Technology Information and KREONET/GLORIAD;
the Polish Ministry of Science and Higher Education and 
the National Science Center;
the Grant of the Russian Federation Government, Agreement No.~14.W03.31.0026; 
the Slovenian Research Agency;
Ikerbasque, Basque Foundation for Science, Spain;
European Union's Horizon 2020 research and innovation programme under grant agreement No. 824093;
the Swiss National Science Foundation; 
the Ministry of Education and the Ministry of Science and Technology of Taiwan;
and the United States Department of Energy and the National Science Foundation.
\end{acknowledgments}


\begin{thebibliography}{13}%
\makeatletter
\providecommand \@ifxundefined [1]{%
 \@ifx{#1\undefined}
}%
\providecommand \@ifnum [1]{%
 \ifnum #1\expandafter \@firstoftwo
 \else \expandafter \@secondoftwo
 \fi
}%
\providecommand \@ifx [1]{%
 \ifx #1\expandafter \@firstoftwo
 \else \expandafter \@secondoftwo
 \fi
}%
\providecommand \natexlab [1]{#1}%
\providecommand \enquote  [1]{``#1''}%
\providecommand \bibnamefont  [1]{#1}%
\providecommand \bibfnamefont [1]{#1}%
\providecommand \citenamefont [1]{#1}%
\providecommand \href@noop [0]{\@secondoftwo}%
\providecommand \href [0]{\begingroup \@sanitize@url \@href}%
\providecommand \@href[1]{\@@startlink{#1}\@@href}%
\providecommand \@@href[1]{\endgroup#1\@@endlink}%
\providecommand \@sanitize@url [0]{\catcode `\\12\catcode `\$12\catcode
  `\&12\catcode `\#12\catcode `\^12\catcode `\_12\catcode `\%12\relax}%
\providecommand \@@startlink[1]{}%
\providecommand \@@endlink[0]{}%
\providecommand \url  [0]{\begingroup\@sanitize@url \@url }%
\providecommand \@url [1]{\endgroup\@href {#1}{\urlprefix }}%
\providecommand \urlprefix  [0]{URL }%
\providecommand \Eprint [0]{\href }%
\providecommand \doibase [0]{http://dx.doi.org/}%
\providecommand \selectlanguage [0]{\@gobble}%
\providecommand \bibinfo  [0]{\@secondoftwo}%
\providecommand \bibfield  [0]{\@secondoftwo}%
\providecommand \translation [1]{[#1]}%
\providecommand \BibitemOpen [0]{}%
\providecommand \bibitemStop [0]{}%
\providecommand \bibitemNoStop [0]{.\EOS\space}%
\providecommand \EOS [0]{\spacefactor3000\relax}%
\providecommand \BibitemShut  [1]{\csname bibitem#1\endcsname}%
\let\auto@bib@innerbib\@empty
\bibitem [{\citenamefont {Collins}(2011)}]{Collins:2011zzd}%
  \BibitemOpen
  \bibfield  {author} {\bibinfo {author} {\bibfnamefont {J.}~\bibnamefont
  {Collins}},\ }\href@noop {} {\bibfield  {journal} {\bibinfo  {journal} {Camb.
  Monogr. Part. Phys. Nucl. Phys. Cosmol.}\ }\textbf {\bibinfo {volume} {32}},\
  \bibinfo {pages} {1} (\bibinfo {year} {2011})}\BibitemShut {NoStop}%
\bibitem [{\citenamefont {Seidl}\ \emph {et~al.}(2015)\citenamefont {Seidl}
  \emph {et~al.}}]{Seidl:2015lla}%
  \BibitemOpen
  \bibfield  {author} {\bibinfo {author} {\bibfnamefont {R.}~\bibnamefont
  {Seidl}} \emph {et~al.} (\bibinfo {collaboration} {Belle Collaboration}),\
  }\href {\doibase 10.1103/PhysRevD.92.092007} {\bibfield  {journal} {\bibinfo
  {journal} {Phys. Rev.}\ }\textbf {\bibinfo {volume} {D 92}},\ \bibinfo
  {pages} {092007} (\bibinfo {year} {2015})},\ \Eprint
  {http://arxiv.org/abs/1509.00563} {arXiv:1509.00563 [hep-ex]} \BibitemShut
  {NoStop}%
\bibitem [{\citenamefont {Altarelli}\ \emph {et~al.}(1979)\citenamefont
  {Altarelli}, \citenamefont {Ellis}, \citenamefont {Martinelli},\ and\
  \citenamefont {Pi}}]{Altarelli:1979kv}%
  \BibitemOpen
  \bibfield  {author} {\bibinfo {author} {\bibfnamefont {G.}~\bibnamefont
  {Altarelli}}, \bibinfo {author} {\bibfnamefont {R.~K.}\ \bibnamefont
  {Ellis}}, \bibinfo {author} {\bibfnamefont {G.}~\bibnamefont {Martinelli}},\
  and\ \bibinfo {author} {\bibfnamefont {S.-Y.}\ \bibnamefont {Pi}},\ }\href
  {\doibase 10.1016/0550-3213(79)90062-2} {\bibfield  {journal} {\bibinfo
  {journal} {Nucl. Phys.}\ }\textbf {\bibinfo {volume} {B160}},\ \bibinfo
  {pages} {301} (\bibinfo {year} {1979})}\BibitemShut {NoStop}%
\bibitem [{\citenamefont {Mulders}\ and\ \citenamefont
  {Van~Hulse}(2019)}]{Mulders:2019mqo}%
  \BibitemOpen
  \bibfield  {author} {\bibinfo {author} {\bibfnamefont {P.~J.}\ \bibnamefont
  {Mulders}}\ and\ \bibinfo {author} {\bibfnamefont {C.}~\bibnamefont
  {Van~Hulse}},\ }\href {\doibase 10.1103/PhysRevD.100.034011} {\bibfield
  {journal} {\bibinfo  {journal} {Phys. Rev.}\ }\textbf {\bibinfo {volume} {D
  100}},\ \bibinfo {pages} {034011} (\bibinfo {year} {2019})},\ \Eprint
  {http://arxiv.org/abs/1903.11467} {arXiv:1903.11467 [hep-ph]} \BibitemShut
  {NoStop}%
\bibitem [{\citenamefont {Leitgab}\ \emph {et~al.}(2013)\citenamefont {Leitgab}
  \emph {et~al.}}]{martin}%
  \BibitemOpen
  \bibfield  {author} {\bibinfo {author} {\bibfnamefont {M.}~\bibnamefont
  {Leitgab}} \emph {et~al.} (\bibinfo {collaboration} {Belle Collaboration}),\
  }\href {\doibase 10.1103/PhysRevLett.111.062002} {\bibfield  {journal}
  {\bibinfo  {journal} {Phys. Rev. Lett.}\ }\textbf {\bibinfo {volume} {111}},\
  \bibinfo {pages} {062002} (\bibinfo {year} {2013})},\ \Eprint
  {http://arxiv.org/abs/1301.6183} {arXiv:1301.6183 [hep-ex]} \BibitemShut
  {NoStop}%
\bibitem [{\citenamefont {Seidl}\ \emph {et~al.}(2017)\citenamefont {Seidl}
  \emph {et~al.}}]{Seidl:2017qhp}%
  \BibitemOpen
  \bibfield  {author} {\bibinfo {author} {\bibfnamefont {R.}~\bibnamefont
  {Seidl}} \emph {et~al.} (\bibinfo {collaboration} {Belle Collaboration}),\
  }\href {\doibase 10.1103/PhysRevD.96.032005} {\bibfield  {journal} {\bibinfo
  {journal} {Phys. Rev.}\ }\textbf {\bibinfo {volume} {D 96}},\ \bibinfo
  {pages} {032005} (\bibinfo {year} {2017})},\ \Eprint
  {http://arxiv.org/abs/1706.08348} {arXiv:1706.08348 [hep-ex]} \BibitemShut
  {NoStop}%
\bibitem [{\citenamefont {Sj{\"o}strand}\ \emph {et~al.}(2006)\citenamefont
  {Sj{\"o}strand}, \citenamefont {Mrenna},\ and\ \citenamefont
  {Skands}}]{Sjostrand:2006za}%
  \BibitemOpen
  \bibfield  {author} {\bibinfo {author} {\bibfnamefont {T.}~\bibnamefont
  {Sj{\"o}strand}}, \bibinfo {author} {\bibfnamefont {S.}~\bibnamefont
  {Mrenna}},\ and\ \bibinfo {author} {\bibfnamefont {P.}~\bibnamefont
  {Skands}},\ }\href {\doibase 10.1088/1126-6708/2006/05/026} {\bibfield
  {journal} {\bibinfo  {journal} {JHEP}\ }\textbf {\bibinfo {volume} {05}},\
  \bibinfo {pages} {026} (\bibinfo {year} {2006})},\ \Eprint
  {http://arxiv.org/abs/hep-ph/0603175} {arXiv:hep-ph/0603175 [hep-ph]}
  \BibitemShut {NoStop}%
\bibitem [{sup()}]{supplement}%
  \BibitemOpen
  \href@noop {} {}\bibinfo {note} {Supplement Material available
  online}\BibitemShut {NoStop}%
\bibitem [{\citenamefont {Seidl}\ \emph {et~al.}(2019)\citenamefont {Seidl}
  \emph {et~al.}}]{Seidl:2019jei}%
  \BibitemOpen
  \bibfield  {author} {\bibinfo {author} {\bibfnamefont {R.}~\bibnamefont
  {Seidl}} \emph {et~al.} (\bibinfo {collaboration} {Belle Collaboration}),\
  }\href {\doibase 10.1103/PhysRevD.99.112006} {\bibfield  {journal} {\bibinfo
  {journal} {Phys. Rev.}\ }\textbf {\bibinfo {volume} {D 99}},\ \bibinfo
  {pages} {112006} (\bibinfo {year} {2019})},\ \Eprint
  {http://arxiv.org/abs/1902.01552} {arXiv:1902.01552 [hep-ex]} \BibitemShut
  {NoStop}%
\bibitem [{\citenamefont {de~Florian}\ \emph {et~al.}(2017)\citenamefont
  {de~Florian}, \citenamefont {Epele}, \citenamefont {Hern{\'a}ndez-Pinto},
  \citenamefont {Sassot},\ and\ \citenamefont {Stratmann}}]{deFlorian:2017lwf}%
  \BibitemOpen
  \bibfield  {author} {\bibinfo {author} {\bibfnamefont {D.}~\bibnamefont
  {de~Florian}}, \bibinfo {author} {\bibfnamefont {M.}~\bibnamefont {Epele}},
  \bibinfo {author} {\bibfnamefont {R.~J.}\ \bibnamefont
  {Hern{\'a}ndez-Pinto}}, \bibinfo {author} {\bibfnamefont {R.}~\bibnamefont
  {Sassot}},\ and\ \bibinfo {author} {\bibfnamefont {M.}~\bibnamefont
  {Stratmann}},\ }\href {\doibase 10.1103/PhysRevD.95.094019} {\bibfield
  {journal} {\bibinfo  {journal} {Phys. Rev.}\ }\textbf {\bibinfo {volume} {D
  95}},\ \bibinfo {pages} {094019} (\bibinfo {year} {2017})},\ \Eprint
  {http://arxiv.org/abs/1702.06353} {arXiv:1702.06353 [hep-ph]} \BibitemShut
  {NoStop}%
\bibitem [{\citenamefont {de~Florian}\ \emph {et~al.}(2015)\citenamefont
  {de~Florian}, \citenamefont {Sassot}, \citenamefont {Epele}, \citenamefont
  {Hern{\'a}ndez-Pinto},\ and\ \citenamefont {Stratmann}}]{deFlorian:2014xna}%
  \BibitemOpen
  \bibfield  {author} {\bibinfo {author} {\bibfnamefont {D.}~\bibnamefont
  {de~Florian}}, \bibinfo {author} {\bibfnamefont {R.}~\bibnamefont {Sassot}},
  \bibinfo {author} {\bibfnamefont {M.}~\bibnamefont {Epele}}, \bibinfo
  {author} {\bibfnamefont {R.~J.}\ \bibnamefont {Hern{\'a}ndez-Pinto}},\ and\
  \bibinfo {author} {\bibfnamefont {M.}~\bibnamefont {Stratmann}},\ }\href
  {\doibase 10.1103/PhysRevD.91.014035} {\bibfield  {journal} {\bibinfo
  {journal} {Phys. Rev.}\ }\textbf {\bibinfo {volume} {D 91}},\ \bibinfo
  {pages} {014035} (\bibinfo {year} {2015})},\ \Eprint
  {http://arxiv.org/abs/1410.6027} {arXiv:1410.6027 [hep-ph]} \BibitemShut
  {NoStop}%
\bibitem [{\citenamefont {Bertone}\ \emph {et~al.}(2018)\citenamefont
  {Bertone}, \citenamefont {Hartland}, \citenamefont {Nocera}, \citenamefont
  {Rojo},\ and\ \citenamefont {Rottoli}}]{Bertone:2018ecm}%
  \BibitemOpen
  \bibfield  {author} {\bibinfo {author} {\bibfnamefont {V.}~\bibnamefont
  {Bertone}}, \bibinfo {author} {\bibfnamefont {N.~P.}\ \bibnamefont
  {Hartland}}, \bibinfo {author} {\bibfnamefont {E.~R.}\ \bibnamefont
  {Nocera}}, \bibinfo {author} {\bibfnamefont {J.}~\bibnamefont {Rojo}},\ and\
  \bibinfo {author} {\bibfnamefont {L.}~\bibnamefont {Rottoli}} (\bibinfo
  {collaboration} {NNPDF collaboration}),\ }\href {\doibase
  10.1140/epjc/s10052-018-6130-4} {\bibfield  {journal} {\bibinfo  {journal}
  {Eur. Phys. J.}\ }\textbf {\bibinfo {volume} {C 78}},\ \bibinfo {pages} {651}
  (\bibinfo {year} {2018})},\ \Eprint {http://arxiv.org/abs/1807.03310}
  {arXiv:1807.03310 [hep-ph]} \BibitemShut {NoStop}%
\bibitem [{\citenamefont {Sato}\ \emph {et~al.}(2016)\citenamefont {Sato},
  \citenamefont {Ethier}, \citenamefont {Melnitchouk}, \citenamefont {Hirai},
  \citenamefont {Kumano},\ and\ \citenamefont {Accardi}}]{Sato:2016wqj}%
  \BibitemOpen
  \bibfield  {author} {\bibinfo {author} {\bibfnamefont {N.}~\bibnamefont
  {Sato}}, \bibinfo {author} {\bibfnamefont {J.~J.}\ \bibnamefont {Ethier}},
  \bibinfo {author} {\bibfnamefont {W.}~\bibnamefont {Melnitchouk}}, \bibinfo
  {author} {\bibfnamefont {M.}~\bibnamefont {Hirai}}, \bibinfo {author}
  {\bibfnamefont {S.}~\bibnamefont {Kumano}},\ and\ \bibinfo {author}
  {\bibfnamefont {A.}~\bibnamefont {Accardi}},\ }\href {\doibase
  10.1103/PhysRevD.94.114004} {\bibfield  {journal} {\bibinfo  {journal} {Phys.
  Rev.}\ }\textbf {\bibinfo {volume} {D 94}},\ \bibinfo {pages} {114004}
  (\bibinfo {year} {2016})},\ \Eprint {http://arxiv.org/abs/1609.00899}
  {arXiv:1609.00899 [hep-ph]} \BibitemShut {NoStop}%
\end{thebibliography}
%
%

\end{document}